\documentclass{PoS}
\usepackage{amsmath,amsfonts,amssymb,mathrsfs}
\usepackage{color}
\usepackage{mathtools} 
\usepackage{graphicx}
\usepackage{caption}
\usepackage{subcaption}
\usepackage{epsfig}
\usepackage{comment}

\newcommand{\mc}[1]{\ensuremath{\mathscr{#1}}}
\newcommand{\mr}[1]{\ensuremath{\mathrm{#1}}}
\newcommand{\val}[1]{\ensuremath{\overline{#1}}}
\newcommand{\sea}[1]{\ensuremath{\underline{#1}}}
\newcommand{\Tr}{\textrm{Tr}}
\newcommand{\bra}[1]{\langle#1|}
\newcommand{\ket}[1]{|#1\rangle}

\title{Flavored pions and kaons at next-to-leading order in mixed-action staggered chiral perturbation theory}

\ShortTitle{Lattice 2013:  Pions and kaons in mixed-action staggered ChPT}

\author{\speaker{Jon A. Bailey}$\,^1$, Jongjeong Kim$^2$, Weonjong Lee$^3$\\
        Lattice Gauge Theory Research Center, CTP, and FPRD, \\
	Department of Physics and Astronomy,
        Seoul National University,
        Seoul, 151-747, South Korea\\
        E-mail: $^1$\email{jabsnu@gmail.com}, 
        $^2$\email{rvanguard@gmail.com}, 
        $^3$\email{wlee@snu.ac.kr}}

\author{Hyung-Jin Kim\\
        Physics Department,
        Brookhaven National Laboratory,
        Upton, NY  11973, USA\\
        E-mail: \email{hjkim@bnl.gov}}

\author{Boram Yoon \\
  Los Alamos National Laboratory,
  Theoretical Division T-2, MS B283\\
  Los Alamos, NM 87545, USA \\
  E-mail: \email{boram@lanl.gov}}

\author{SWME Collaboration\thanks{W.~Lee is supported by the Creative
    Research Initiatives program (2013-003454) of the NRF grant funded
    by the Korean government (MSIP).  Brookhaven National Laboratory
    is operated under Contract No.~DE-AC02-98CH10886 with the
    U.S. Department of Energy.  J.A.B. is supported by the Basic
    Science Research Program (2013009149) of the National Research
    Foundation of Korea (NRF) funded by the Ministry of Education.}}

\abstract{ Different versions of improved staggered fermions can be
  used as valence quarks to reduce discretization effects in lattice
  QCD calculations while increasing statistics on existing staggered
  gauge ensembles.  Such mixed-action simulations can be used to
  improve determinations of light quark masses, Gasser-Leutwyler
  couplings, decay constants, and other parameters relevant to
  particle phenomenology.  We recall the generalization of ordinary,
  unmixed staggered chiral perturbation theory required to describe
  data from lattice calculations with a mixed action such as with HYP
  staggered valence quarks and asqtad sea quarks.  We calculate the
  next-to-leading order loop diagrams contributing to the masses and
  decay constants of the flavored pseudo-Goldstone bosons of all
  tastes and here report results for the decay constants and
  valence-valence masses.  }

\FullConference{31$^{st}$ International Symposium on Lattice Field
  Theory --- LATTICE 2013\\ July 29 -- August 3, 2013\\ Mainz,
  Germany}

\begin{document}

\section{Introduction}
Staggered chiral perturbation theory (SChPT) has been used extensively
to control extrapolations from unphysical light-quark simulation
masses to physical masses and to remove dominant light quark and gluon
discretization effects~\cite{ref:SChPT}.  Mixed-action ChPT was
developed for simulations performed with Ginsparg-Wilson valence
quarks and (less computationally expensive) Wilson sea
quarks~\cite{Bar:first}.  The formalism for staggered sea quarks and
Ginsparg-Wilson valence quarks was developed in
Ref.~\cite{Bar:2005tu}.  Mixed-action ChPT for differently improved
staggered fermions was introduced for calculations of the
$K^0-\overline{K^0}$ bag parameter ~\cite{Bae:2010ki} and the
$K\to\pi\ell\nu$ vector form factor~\cite{Bazavov:2012cd}.  As a
by-product, the pseudo-Goldstone boson (PGB) masses and propagators
were calculated at tree level in mixed-action SChPT.  While
contributions to overall errors were very small, some of the
low-energy couplings (LECs) of mixed-action SChPT were not well
determined by the data~\cite{Bae:2010ki,Bazavov:2012cd}.

Here we present a calculation of the next-to-leading order (NLO) loop
corrections to the masses and decay constants of the flavor-charged
PGBs in all taste irreps.  Analyses of corresponding spectrum data may
improve our knowledge of LECs that are poorly determined by existing
simulations.  Our results can also be used to improve determinations
of the light quark masses, Gasser-Leutwyler couplings, and pion and
kaon decay constants.  Data for such analyses could be generated with,
{\it e.g.}, asqtad sea quarks and valence HYP, or with HISQ sea quarks
and improved staggered valence quarks that would allow for
simulating relativistic bottom quarks at lattice spacings $\gtrsim
0.03\ \mr{fm}$.  Mixed-action SChPT results could facilitate future
calculations of quantities such as $|V_{cb}|$ and the $K\to\pi\pi$
amplitudes.

%
%

\section{\label{sec:mixed-SChPT} Mixed-action staggered chiral perturbation theory}
Mixed-action theories are generalizations of partially quenched
theories with different valence and sea quark actions.  The symmetries
relating valence and sea quarks are broken, but with differently
improved versions of the same action, the symmetries of the valence
(sea) sector are the same as in the unmixed theory.  As for ordinary,
unmixed SChPT, mixed-action SChPT is constructed in two steps.  First
one builds the Symanzik effective theory (SET) for the (mixed-action)
lattice theory.  One then maps the operators of the SET into those of
ChPT~\cite{CBnotes}.
\subsection{The leading order Lagrangian}
Mapping the SET Lagrangian into the chiral theory through
NLO, we have
\begin{align}
  S_\mr{eff} = S_\mr{QCD} + a^2 S_6 + \dots \longrightarrow 
  \mc{L}_\mr{SChPT} = 
  \mc{L}_\mr{LO,ChPT} + a^2 \mc{V} + \dots \,.
\end{align}
Here $S_\mr{QCD}$ ($\mc{L}_\mr{LO,ChPT}$) has the form of the QCD action (the
leading order Lagrangian of continuum ChPT), but respects the doubler symmetry,
taste SU(4).  As in continuum ChPT, the Lagrangian $\mc{L}_\mr{LO,ChPT}$
contains kinetic energy, mass, and anomaly terms.  The operators in $S_6$ break
the continuum symmetries, including those relating valence and sea quarks, to
those of the mixed-action lattice theory.  A subset of four-fermion operators
in $S_6$ respects $\Gamma_4\rtimes\,$SO(4) $\subset$ SU(4).  They map to the
potential $\mc{V}$ and can be obtained from those of the unmixed SET by
introducing projection operators $P_{v,\sigma}$ onto the valence and sea
sectors and allowing the LECs to differ in the valence and sea sectors.
Generically, we have
\begin{align}
c\,\bar{\psi}(\gamma_s\otimes\xi_t)\psi\,\bar{\psi}(\gamma_s\otimes\xi_t)\psi
\longrightarrow\,c_{vv}\,&\bar{\psi}(\gamma_s\otimes\xi_t)P_v\psi\,\bar{\psi}(\gamma_s\otimes\xi_t)P_v\psi + (v\rightarrow\sigma) \\
+\,2c_{v\sigma}\,&\bar{\psi}(\gamma_s\otimes\xi_t)P_v\psi\,\bar{\psi}(\gamma_s\otimes\xi_t)P_\sigma\psi\nonumber
\end{align}
where $\gamma_s$ ($\xi_t$) is a spin (taste) matrix.  To construct the
potential $\mc{V}$, the projection operators are conveniently included in 
spurions.  The resulting potential is $\mc{V} = \mc{U} + \mc{U}^\prime -
C_\mr{mix}\Tr(\tau_3 \Sigma \tau_3 \Sigma^\dag)$, where $\mc{U}$
($\mc{U}^\prime$) contains single-(double-)trace operators that are direct
generalizations of those in unmixed SChPT, and the last term is a taste-singlet
potential new in the mixed-action theory, with $\tau_3\equiv P_\sigma - P_v$.
The operators in $\mc{U}^{(\prime)}$ have independent LECs for the
valence-valence, sea-sea, and valence-sea sectors.  For example,
\begin{align}
C_1\Tr(\xi_5\Sigma\xi_5\Sigma^{\dagger})
\longrightarrow C^{vv}_1\Tr(\xi_5P_v\Sigma\xi_5P_v\Sigma^{\dagger}) + (v\rightarrow\sigma)
+\,C^{v\sigma}_1&[\Tr(\xi_5P_v\Sigma\xi_5P_\sigma\Sigma^{\dagger})+ p.c.]\,.
\end{align}
where $p.c.$ means parity conjugate.
For the unmixed case, $C_1^{vv}=C_1^{\sigma\sigma}=C_1^{v\sigma}=C_1$,
$C_\mr{mix}=0$, and the potential reduces to that of ordinary SChPT.
The full expressions for the potentials $\mc{U}^{(\prime)}$ are
somewhat lengthy, and we defer writing them down~\cite{SWME:2013}.
\subsection{Tree-level propagators and flavored PGB masses}
The potential $\mc{V}$ contributes to the tree-level masses of the
PGBs, which fall into irreps of $\Gamma_4\rtimes\,$SO(4).  For a taste
$t$ PGB $\phi^t_{xy}$ composed of quarks with flavors $x,y,\ x\neq y$,
\begin{align}
m_{xy,\,t}^2=\mu (m_x + m_y) + a^2 \Delta_F^{xy},\quad t\in F\in\{P,A,T,V,I\} \,,
\label{eq:tree}
\end{align}
where $F$ labels the taste $\Gamma_4\rtimes\,$SO(4) irreps
(pseudoscalar, axial, tensor, vector, or scalar), and $\mu$ is the
condensate parameter.  $\Delta_F^{xy}$ is the tree-level mass
splitting, which depends on the LECs in $\mc{U}$ and $C_\mr{mix}$, as
well as the sector (valence or sea) of the flavors $x,y$.  We have
\begin{align}
\Delta^{vv}_a = &\frac{8}{f^2}\sum_{b\neq I}\,C^{vv}_b\,(1-\theta^{ab}\theta^{b5}),\quad
\Delta^{\sigma\sigma}_a = \frac{8}{f^2}\sum_{b\neq I}\,C^{\sigma\sigma}_b\,(1-\theta^{ab}\theta^{b5})
\label{eq:mass_split_ee}\\
\Delta^{v\sigma}_a&=\frac{16C_\mr{mix}}{f^2}+\frac{8}{f^2}\sum_{b\neq I}\left[\tfrac{1}{2}(C^{vv}_b+C^{\sigma\sigma}_b)-C^{v\sigma}_b\theta^{ab}\theta^{b5}\right]\,,
\label{eq:mass_split_vs}
\end{align}
where the splitting is $\Delta_a^{vv(\sigma\sigma)}$ if both quarks
are valence (sea) quarks and $\Delta_a^{v\sigma}$ otherwise;
sub(super)scripts $a, b$ are taste indices labeling the generators of
the fundamental irrep of U(4); and $\theta^{ab}=+1(-1)$ if the
generators for $a$ and $b$ (anti)commute.  
The LECs $C_b^{vv, \sigma\sigma, v\sigma}$ come from the potential
$\mathcal{U}$ and are defined in analogy with the unmixed case in
Ref.~\cite{SWME:2011aa}.
%
%
The residual chiral symmetry in the valence-valence sector, as for the
unmixed theory (equivalently, the sea-sea sector), implies $F=P$
particles are Goldstone bosons for $a\neq 0,\ m_q = 0$, and therefore
$\Delta_P^{vv}=0$.  The same is not true for the taste pseudoscalar,
valence-sea PGBs, and generically, $\Delta_P^{v\sigma}\neq 0$.

In the flavor-neutral sector, $x = y$, the PGBs mix in the taste
singlet, vector, and axial irreps.  The Lagrangian mixing terms
(hairpin terms) are
\begin{align}
\tfrac{1}{2}\,\delta\,\phi^I_{ii}\phi^I_{jj} + \tfrac{1}{2}\,\delta_V^{vv}\,\phi^\mu_{\val{i}\val{i}}\phi^\mu_{\val{j}\val{j}} + \tfrac{1}{2}\,\delta_V^{\sigma\sigma}\,\phi^\mu_{\sea{i}\sea{i}}\phi^\mu_{\sea{j}\sea{j}} + \delta_V^{v\sigma}\,\phi^\mu_{\val{i}\val{i}}\phi^\mu_{\sea{j}\sea{j}} + (V\rightarrow A,\ \mu \rightarrow \mu5)\,, \label{eq:hair}
\end{align}
where $i,j$ are flavor indices; $\mu$ ($\mu5$) is a taste index in the
vector (axial) irrep; and we use an overbar (underbar) to restrict
summation to the valence (sea) sector.
The $\delta$-term comes from the anomaly contribution.  In continuum
ChPT, taking $\delta\rightarrow\infty$ at the end of the calculation
decouples the $\eta^\prime$~\cite{Sharpe:2001fh}.  In SChPT, taking
$\delta\rightarrow\infty$ decouples the $\eta^\prime_I$.  The
flavor-singlets in other taste irreps are PGBs and do not
decouple~\cite{Aubin:2003mg}.  The
$\delta_{V,\,A}^{vv,\sigma\sigma,v\sigma}$-terms are lattice artifacts
from the potential $a^2\mathcal{U}^\prime$, and the couplings
$\delta_{V,\,A}^{vv,\sigma\sigma,v\sigma}$ depend linearly on its
LECs.

Although the mass splittings and hairpin couplings are different in the three sectors, we find the tree-level
propagator can be written in the same form as in the unmixed case.  We have
\begin{align}
G^{ab}_{ij,\,kl}(p^2)=\delta^{ab}\left(\frac{\delta_{il}\delta_{jk}}{p^2+\mu(m_{i}+m_{j})+a^2\Delta^{ij}_a}+\delta_{ij}\delta_{kl}\,D^a_{il}\right),
\label{treeprop}
\end{align}
where the disconnected propagators vanish by definition in the pseudoscalar and tensor irreps, and for the singlet, vector, and axial irreps,
\begin{align}
D^a_{ij} \equiv \frac{-1}{I_a J_a}
\begin{dcases}\label{Dprop}
\frac{\delta^{ij}_a}{1+\delta^{\sigma\sigma}_a\sea{\sigma_a}} & \text{$ij\notin vv$}\\
\left(\frac{(\delta^{v\sigma}_a)^2/\delta^{\sigma\sigma}_a}{1+\delta^{\sigma\sigma}_a\sea{\sigma_a}}+
\delta^{vv}_a-(\delta^{v\sigma}_a)^2/
\delta^{\sigma\sigma}_a\right)
& \text{$ij\in vv$,}
\end{dcases}
\end{align}
where $\delta_I^{ij}\equiv \delta$, $I_a \equiv p^2+2\mu
m_{i}+a^2\Delta^{ii}_a$, $J_a \equiv p^2+2\mu m_{j}+a^2\Delta^{jj}_a$,
and we used the replica method to quench the valence
quarks~\cite{Damgaard:2000gh} and root the sea
quarks~\cite{Aubin:2003mg}, so that
\begin{align}
\sea{\sigma_a} & \equiv \sum_{\sea{i}}\frac{1}{p^2 + 2\mu m_{\sea{i}} + a^2 \Delta_a^{\sigma\sigma}}=\tfrac{1}{4}\sum_{\sea{i}^\prime}\frac{1}{p^2 + 2\mu m_{\sea{i}^\prime} + a^2 \Delta_a^{\sigma\sigma}}
\\
\val{\sigma_a} & \equiv \sum_{\val{i}}\frac{1}{p^2 + 2\mu m_{\val{i}} + a^2 \Delta_a^{vv}}=0 \,.
\end{align}
The index $\sea{i}$ runs over the replica flavors which include the
taste degrees of freedom.  The index $\sea{i}^\prime$ is summed over
the physical sea quark flavors such as $u,d,s$.
%
%
\begin{figure}[tbph]
\begin{center}
\begin{subfigure}[b]{0.17\textwidth}
\includegraphics[width=\textwidth]{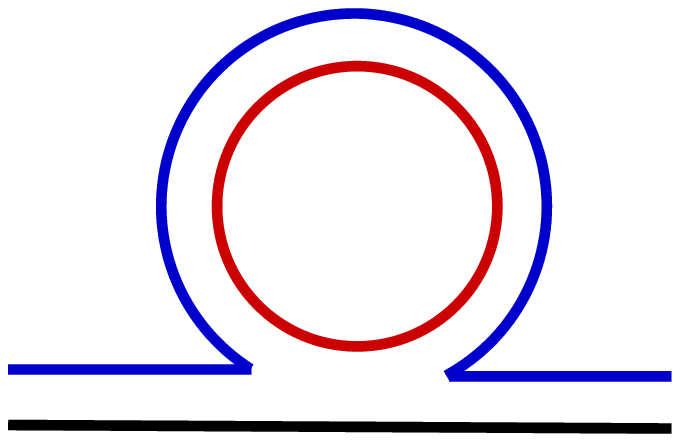}
\caption{}
\label{fig:qflow-a}
\end{subfigure}
\
\begin{subfigure}[b]{0.17\textwidth}
\includegraphics[width=\textwidth]{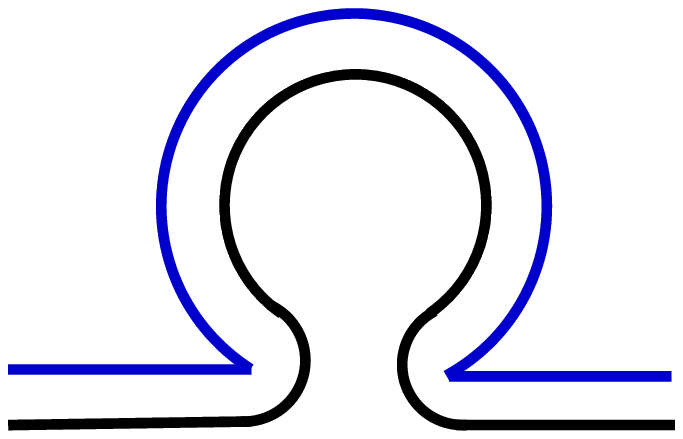}
\caption{}
\label{fig:qflow-b}
\end{subfigure}
\ 
\begin{subfigure}[b]{0.17\textwidth}
\includegraphics[width=\textwidth]{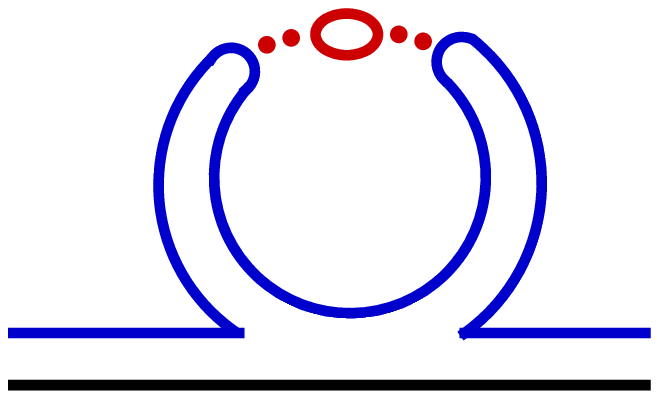}
\caption{}
\label{fig:qflow-c}
\end{subfigure}
\
\begin{subfigure}[b]{0.17\textwidth}
\includegraphics[width=\textwidth]{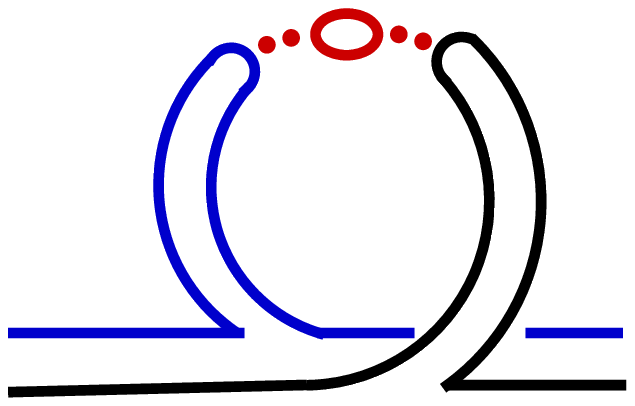}
\caption{}
\label{fig:qflow-d}
\end{subfigure}%
\ 
\begin{subfigure}[b]{0.17\textwidth}
\includegraphics[width=\textwidth]{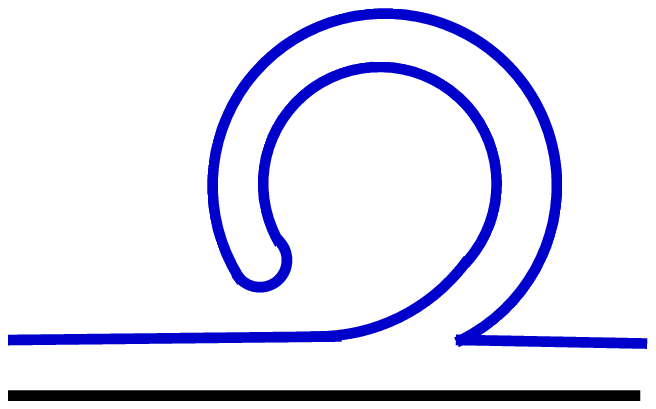}
\caption{}
\label{fig:qflow-e}
\end{subfigure}
\
\begin{subfigure}[b]{0.17\textwidth}
\includegraphics[width=\textwidth]{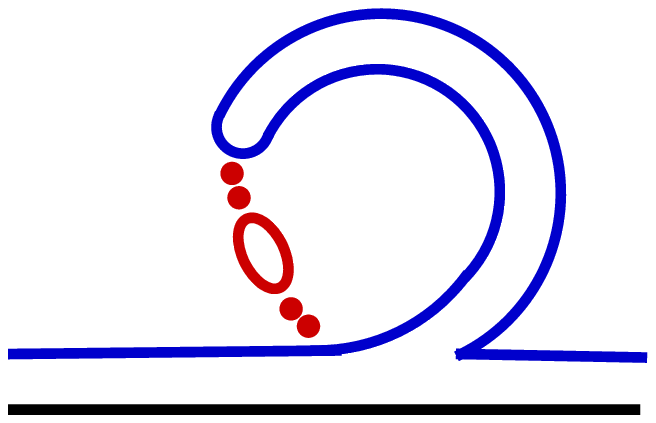}
\caption{}
\label{fig:qflow-f}
\end{subfigure}
\
\begin{subfigure}[b]{0.2\textwidth}
\includegraphics[width=\textwidth]{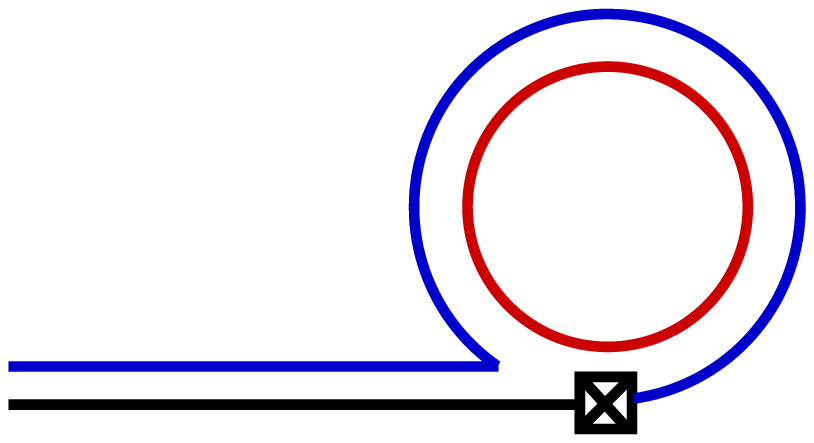}
\caption{}
\label{fig:qflow-g}
\end{subfigure}%
\ 
\begin{subfigure}[b]{0.2\textwidth}
\includegraphics[width=\textwidth]{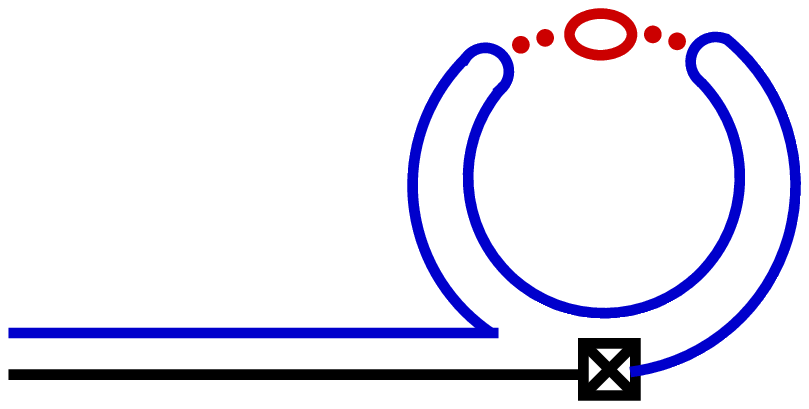}
\caption{}
\label{fig:qflow-h}
\end{subfigure}
\
\begin{subfigure}[b]{0.2\textwidth}
\includegraphics[width=\textwidth]{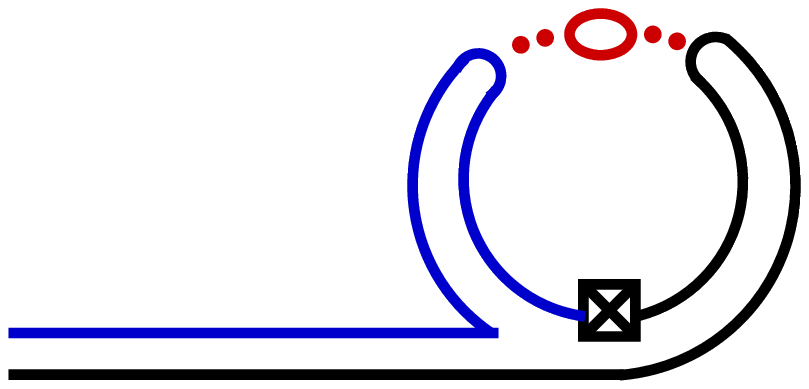}
\caption{}
\label{fig:qflow-i}
\end{subfigure}
\caption{Quark flows for the NLO self-energy tadpoles (a-f) and
  current-vertex loops (g-i).  The $x$ and $y$ quarks are represented
  by blue and black lines, sea quarks are represented by red lines,
  and current insertions are represented by crossed boxes.}
\label{fig:qflowdiag} 
\end{center} 
\end{figure}
\section{\label{sec:self-energy}NLO loop corrections to masses}
For a taste $t$ PGB $\phi^t_{xy}$ composed of quarks with flavors
$x,y,\ x\neq y$, the mass is defined in terms of the self-energy, as
in continuum ChPT.  The NLO mass can be obtained by adding the NLO
self-energy to the tree-level value,
\begin{align}
  M_{xy,\,t}^2 = m_{xy,\,t}^2 + \Sigma_{xy,\, t}(-m_{xy,\,t}^2) \,.
\end{align}
$\Sigma_{xy,\, t}$ consists of connected and disconnected tadpole
loops with vertices from the LO Lagrangian at $\mc{O}(\phi^4)$ and
analytic terms with vertices from the NLO Lagrangian at
$\mc{O}(\phi^2)$.

For a general $\Gamma_4\rtimes\,$SO(4) irrep, the calculation of the
valence-valence, flavored PGB self-energies proceeds as for the
unmixed case.  Quark flows for the tadpoles are shown in graphs (a-f)
of Fig.~\ref{fig:qflowdiag}.  The kinetic energy, mass, and $\mc{U}$
vertices yield graphs of type (a), (c), and (d), and the taste-singlet
potential vertices ($\propto C_\mr{mix}$) yield graphs of type (a),
\begin{align}
\frac{a^2C_\mr{mix}}{3f^2(4\pi f)^2}\,\sum_{\val{i}^\prime\sea{i}^\prime a}\,\ell(m^2_{\val{i}^\prime \sea{i}^\prime,\,a})\,,\label{eq:Cmix}
\end{align}
where $\val{i}^\prime$ is summed over $x,y$; $\sea{i}^\prime$ is summed over
the physical sea quarks; and $\ell(m^2)\equiv m^2 \ln (m^2/\Lambda^2)$ is the
chiral logarithm, with $\Lambda$ the scale of dimensional regularization.  

Vertices from $\mc{U}^\prime$ yield graphs of type (b), (e), and (f).
The hairpin vertex graphs are of type (e) and (f).  As in the unmixed
case, they ((e) and (f)) can be combined and converted into the form
of (d).  In the mixed-action case, the necessary identity is
\begin{align}
\frac{\delta^{vv}_a}{p^2+2\mu m_x + a^2\Delta^{vv}_a} + \frac{\delta^{v\sigma}_a}{4}\sum_{\sea{i}^\prime} D^a_{x\sea{i}^\prime} = -(p^2+2\mu m_y + a^2\Delta^{vv}_a)D^a_{xy},\quad a\in V,A, \label{eq:coolID}
\end{align}
where $x,y$ are valence quarks.  This relation follows from
Eq.~\eqref{Dprop}.
Graphs of type (b) come from vertices $\propto \omega^{vv}_{V,A}$;
they have the same form as those in the unmixed
case~\cite{SWME:2011aa}.

Adding the graphs and evaluating the result at
$p^2=-m^2_{xy,\,t}=-\mu(m_x+m_y)-a^2\Delta^{vv}_t$, we have the NLO,
one-loop contributions to the self-energies of the valence-valence,
flavored PGBs,
\begin{align}
-\Sigma^{\text{NLO loop}}_{xy,\,t}(-m^2_{xy,\,t})=
\frac{a^2}{48(4\pi f)^2}\,\sum_a\Biggl[&\left(\Delta^\mr{mix}_{at}-\Delta^{vv}_t-\Delta^{v\sigma}_a+\frac{16C_\mr{mix}}{f^2}\right)\,\sum_{\val{i}^\prime\sea{i}^\prime} \ell(m^2_{\val{i}^\prime\sea{i}^\prime,\,a})\label{eq:lQ}\\
&+\ \frac{3}{2}\Biggl(\sum_{b\in V,A}\omega^{vv}_b\tau_{abt}\tau_{abt}(1+\theta^{ab})\Biggr)\ell(m^2_{xy,\,a})\Biggr]\\
+\ \frac{1}{12(4\pi f)^2}\,\sum_a\Biggl[&a^2\left(\Delta^{vv}_{at}-\Delta^{vv}_t-\Delta^{vv}_a\right)\int\frac{d^4q}{\pi^2}(D_{xx}^a+D_{yy}^a)\\
+ \int\frac{d^4q}{\pi^2}&\biggl[\Bigl(2(1-\theta^{at}) + \rho^{at}\Bigr)q^2
+\Bigl(2(1+2\theta^{at}) + \rho^{at}\Bigr)m^2_{xy,\,5}\\
&+\ 2a^2\Delta^{\prime vv}_{at} + a^2\Bigl(2\theta^{at}\Delta^{vv}_t + (2+\rho^{at})\Delta^{vv}_a\Bigr)\biggr]D_{xy}^a\Biggr]\,,\label{eq:Dxy}
\end{align}
where $\rho^{at}\equiv -4(2+\theta^{at})$ unless $a=I$, when it vanishes, $\tau_{abt}\equiv\Tr(T^aT^bT^t)$ is a trace over (a product of) generators of U(4), $\omega^{vv}_b$ depends on the LECs of $\mc{U}^\prime$, and 
\begin{align}
\Delta^{vv}_{at}&\equiv\frac{8}{f^2}\sum_{b\neq I} C^{vv}_b(5+3\theta^{ab}\theta^{bt}-4\theta^{5b}\theta^{bt}-4\theta^{ab}\theta^{b5})\,,\\
\Delta^{\prime vv}_{at}&\equiv\frac{8\theta^{at}}{f^2}\sum_{b\neq I} C^{vv}_b(1+3\theta^{ab}\theta^{bt}-2\theta^{5b}\theta^{bt}-2\theta^{ab}\theta^{b5})\,,\\
\Delta^\mr{mix}_{at}&\equiv \frac{8}{f^2}\sum_{b\neq I}\left[\tfrac{1}{2}(9C^{vv}_b + C^{\sigma\sigma}_b) + C^{v\sigma}_b(3\theta^{ab}\theta^{bt}-4\theta^{ab}\theta^{b5})-4C^{vv}_b\theta^{5b}\theta^{bt}\right]\,.
\end{align}
The form of Eqs.~\eqref{eq:lQ}-\eqref{eq:Dxy} is the same as that in ordinary
SChPT~\cite{SWME:2011aa}; the differences are in the definition of the
disconnected propagators and the LECs of the effective field theory.

The valence-valence, taste-pseudoscalar PGBs are true Goldstone bosons in the
chiral limit, $m_x,m_y\rightarrow 0,\ a\neq 0$.  Setting $t=5$ in Eqs.~\eqref{eq:lQ}
through \eqref{eq:Dxy} gives
\begin{align}
-\Sigma^{\text{NLO loop}}_{xy,\,5}(-m^2_{xy,\,5})=\frac{\mu(m_x+m_y)}{2(4\pi f)^2}\,\sum_a\theta^{a5}\int\frac{d^4q}{\pi^2}D^a_{xy}\,,
\end{align}
which is the generalization of the results of Ref.~\cite{Aubin:2003mg} to the mixed-action case.  Only graphs of type (d) contribute.

For the valence-sea, flavored PGB self-energies, the calculation proceeds
similarly, but there is no symmetry under $x \leftrightarrow y$, and the taste
pseudoscalars are not Goldstone bosons in the chiral limit.  Graphs of type (a)
contribute chiral logarithms from valence-sea and sea-sea PGBs in the loop, and
graphs of type (b) enter with valence-sea PGBs in the loop.  The elimination of
contributions of type (e) and (f) in favor of contributions of type (d) leaves
leftover chiral logarithms of the valence-valence PGBs,
$\ell(m^2_{\val{x}\val{x},\,a})$, multiplied by combinations of hairpin
couplings that vanish in the unmixed limit.  For $a\neq 0,\ m_x,m_y\rightarrow
0$, graphs of type (a), (b), (c), and (e) contribute; the valence-sea,
taste-pseudoscalars PGBs are not true Goldstone bosons in the chiral limit.
However, no new loop integrals arise.  The details of the calculation and the
results will be presented in Ref.~\cite{SWME:2013}.

\section{\label{sec:decay}NLO loop corrections to decay constants}

The decay constants are defined in terms of the matrix elements of the axial
currents,
\begin{align}
  -i f_{xy,\,t}\, p_\mu = \bra{0}\, j^{\mu5}_{xy,\,t}\, \ket{\phi^t_{xy}(p)} \,.
\end{align}
The NLO corrections come from wave function renormalization [graphs (a), (c),
and (d) of Fig.~\ref{fig:qflowdiag}], insertions of the $\mc{O}(\phi^3)$-terms
of the LO current [graphs (g), (h), and (i) of Fig.~\ref{fig:qflowdiag}], and
NLO analytic terms~\cite{Aubin:2003uc}.  The LO current is determined by the
kinetic energy vertices of the LO Lagrangian, and is therefore the same as in
unmixed SChPT.  Likewise, the NLO wave function renormalization corrections are
determined by self-energy contributions from tadpoles with kinetic energy
vertices from the LO Lagrangian.  To generalize the results of the unmixed
case, we have only to replace the propagators with those of the mixed-action
theory; nothing in the calculation of the relevant part of the self-energies or
the current-vertex loops is sensitive to the sector of the external quarks.  We
have
\begin{align}
\frac{f^\text{NLO loop}_{xy,\,t}}{f} = 1 - \frac{1}{8(4\pi f)^2}\sum_a\left[\frac{1}{4}\sum_{\val{i}^\prime\sea{i}^\prime} \ell(m^2_{\val{i}^\prime\sea{i}^\prime,\,a}) + \int\frac{d^4q}{\pi^2}(D^a_{xx}+D^a_{yy}-2\theta^{at}D^a_{xy})\right]\,.\label{eq:decay_fin}
\end{align}
where $x,y$ can take either valence ($\val{x},\val{y}$) or sea 
($\sea{x},\sea{y}$) indices.
This result holds for valence-valence ($\val{x} \neq \val{y}$),
sea-sea ($\sea{x} \neq \sea{y}$), and valence-sea ($\val{x}$ and
$\sea{y}$) PGBs, and has the same form as in the unmixed
theory~\cite{Bailey:2012jy}.

\section{\label{sec:sum}Summary}

Using mixed-action SChPT, we have calculated the NLO loop corrections
to the valence-valence and valence-sea masses and decay constants of
flavored PGBs in all taste irreps.  The results have been
cross-checked by performing two independent calculations, and we have
verified the results reduce correctly when valence and sea quark
actions are the same.  In the valence-valence sector, the taste
pseudoscalars are Goldstone bosons in the chiral limit, at nonzero
lattice spacing.  The NLO analytic terms arise from tree-level
contributions of the (NLO) Gasser-Leutwyler and generalized Sharpe-Van
de Water Lagrangians.  They have the same form as in the unmixed case,
with different couplings in the valence-valence, sea-sea, and
valence-sea sectors.

The NLO loop corrections to the self-energies of the valence-valence,
flavored PGBs are given in Eqs.~\eqref{eq:lQ}-\eqref{eq:Dxy}; those
for the valence-valence and valence-sea decay constants are given in
Eq.~\eqref{eq:decay_fin}.  They have the same form as the results in
ordinary, unmixed SChPT.  The results for the valence-sea
self-energies will be included in Ref.~\cite{SWME:2013}.

\end{document}